\documentclass[12pt]{article}
\usepackage{epsfig}
\usepackage{rotating}
\usepackage{booktabs}
\usepackage{lscape}
\usepackage{amsmath}
\usepackage{multirow}
\usepackage[usenames]{color}

\usepackage{color}

\usepackage{cite}

\begin{document}

\begin{center}

\vspace*{1.0cm}

{\Large \bf{First direct search for $2\epsilon$ and $\epsilon\beta^+$ decay of $^{144}$Sm
and $2\beta^-$ decay of $^{154}$Sm}}

\vskip 1.0cm

{\bf P.~Belli$^{a,b}$, R.~Bernabei$^{a,b,}$\footnote{Corresponding
author. {\it E-mail address:} rita.bernabei@roma2.infn.it.}, 
R.~S.~Boiko$^{c,d}$, F.~Cappella$^{e,f}$,\\
V.~Caracciolo$^{a,b}$, R.~Cerulli$^{a,b}$, F.~A.~Danevich$^c$, A. Di Marco$^{a,b}$,\\
A. Incicchitti$^{e,f}$, B.~N.~Kropivyansky$^c$, M.~Laubenstein$^g$, S.~Nisi$^g$, \\
D.~V.~Poda$^{c,h}$, O.~G.~Polischuk$^{c,b}$ and V.~I.~Tretyak$^c$}

\end{center}

\vskip 0.3cm

{\footnotesize
\noindent $^a$ INFN sezione Roma ``Tor Vergata'', I-00133 Rome, Italy\\
$^b$ Dipartimento di Fisica, Universit\`a di Roma ``Tor Vergata'', I-00133, Rome, Italy\\
$^c$ Institute for Nuclear Research, National Academy of Sciences of 
Ukraine, 03028 Kyiv, Ukraine\\
$^d$ National University of Life and Environmental Sciences of Ukraine, 03041 Kyiv, Ukraine\\
$^e$ INFN sezione Roma, I-00185 Rome, Italy\\
$^f$ Dipartimento di Fisica, Universit\`a di Roma ``La Sapienza'', I-00185 Rome, Italy\\
$^g$ INFN, Laboratori Nazionali del Gran Sasso, I-67100 Assergi (AQ), Italy\\
$^h$ CSNSM, Univ. Paris-Sud, CNRS/IN2P3, Universit\'e Paris-Saclay, 91405 
Orsay, France}

\vskip 0.5cm

\begin{abstract}

The first direct search for the double electron capture ($2\epsilon$) and the electron capture with
positron emission ($\epsilon\beta^+$) in $^{144}$Sm to the ground state and to the excited
levels of $^{144}$Nd was realized by measuring -- over 1899 h -- a 342 
g sample of highly purified samarium oxide 
(Sm$_2$O$_3$) with the ultra-low background HP-Ge $\gamma$
spectrometer GeCris (465 cm$^3$) at the STELLA facility of the Gran Sasso National 
Laboratory (LNGS). No effect was observed and half-life limits were estimated at the level of
$T_{1/2} \sim (0.1-1.3) \times 10^{20}$ yr (90\% C.L.).
Moreover, for the first time half-life limits of the double beta ($2\beta^-$) decay of  $^{154}$Sm to several excited
levels of $^{154}$Gd have been set; they are at the level of $T_{1/2} \sim (0.06-8) \times 
10^{20}$ yr (90\% C.L.).
\end{abstract}

\vskip 0.4cm


\section{Introduction}

The studies on neutrino oscillations provide information about differences of squared neutrino 
masses, and show that the neutrino 
mass matrix of weak neutrino eigenstates is non-diagonal. In this theoretical framework, processes which 
can occur in the standard model (SM) are modified
and other processes due to non vanishing neutrino masses can occur.

Double beta decay with neutrino emission conserves the lepton number, providing a confirmation of the 
SM of weak 
interaction. On the contrary, the neutrino-less double beta  decay (0$\nu$2$\beta$), which violates the 
lepton number by two units, can be a signature of new physics beyond the SM. Thus, investigations 
on neutrino-less double beta decay can provide information about neutrino properties, weak interaction, lepton number 
violation, and outline new theoretical frameworks \cite{barea,Rodejo,deppis,bilenky,oro,vergados}.
While the two neutrino  mode of double beta decay is already observed in several nuclides with the 
half-lives $T^{2\nu2\beta}_{1/2} \sim 10^{18} - 10^{24}$ yr, for the $0\nu2\beta$ decay 
the work is in progress. Typically, sensitivities on the half-life of the $0\nu2\beta$ decay up to 
$T^{0\nu2\beta}_{1/2} \ge 10^{24} - 10^{26}$ yr  have been  
published with various kinds of experimental set-ups applying different procedures
\cite{oro,tretyak,elliot,giuliani,cremonesi,sara,arnold,gando,albert,alduino,Aa,agostini,azz}. 
Much more modest sensitivities have been presented in the investigations of the double beta plus 
processes ($2\epsilon$, $\epsilon\beta^+$ and $2\beta^+$)
\cite{tretyak,maala}. This could be ascribed e.g. to: (i) the technological progress  in making 
detectors containing 
2$\beta^-$ emitters, as e.g. germanium detectors; (ii)  the typically larger abundance of the most 
promising 2$\beta^-$ emitters with respect to the 2$\beta^+$ 
emitters; (iii) the higher decay probability of 2$\beta^-$ processes with respect to 
that of the 2$\beta^+$ decay processes, which suffer of the suppression effect due to a smaller 
phase space factor. 
However, the investigations of $\varepsilon\beta^+$ and $2\beta^+$ processes 
strongly contribute to clarify the mechanism of the 
$0\nu2\beta^-$ decay \cite{hirsch}. Moreover, investigations of the 
0$\nu2\epsilon$ processes are also supported by   
the possibility of a resonant enhancement of the capture rate because of the  mass degeneracy between the initial and 
the final nucleus \cite{reso,reso1,reso2,reso3,Krivo,Geo,Eli}.

Experimental searches for $2\nu2\beta$ processes, which are allowed in the SM and are already observed in 
dozen of nuclides \cite{Saak} but not in Sm isotopes, give possibilities to check 
aspects of the theoretical calculations 
which are used also in the estimates of the half-lives of the $0\nu2\beta$ mode.

\begin{figure}[!ht]
  \centering
  \includegraphics[width=\textwidth]{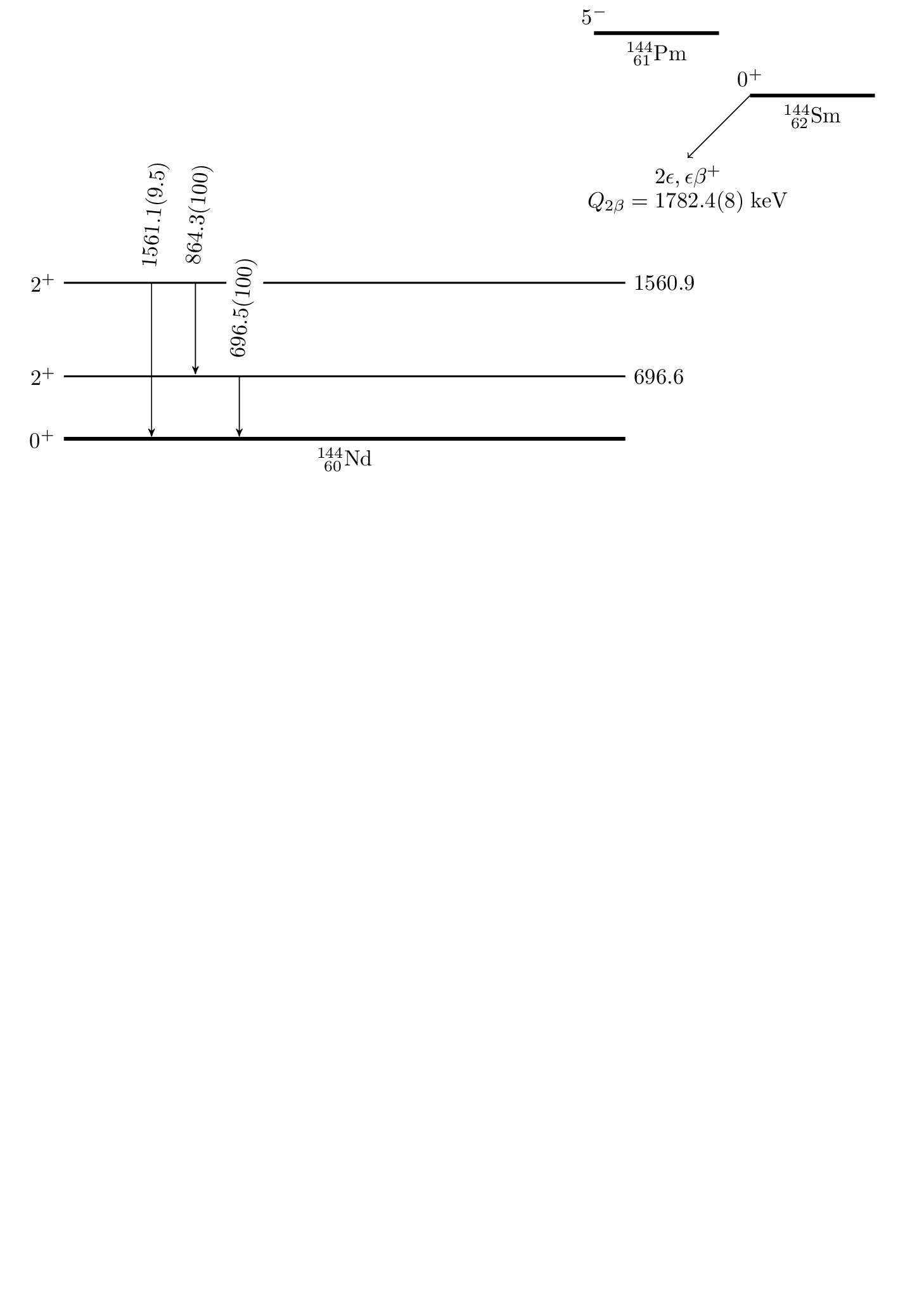}
\caption{Simplified scheme of the double beta decay of $^{144}$Sm \cite{tabNd144}. The energies of the excited levels and
of the emitted $\gamma$ quanta are in keV; the relative intensities of the $\gamma$ quanta
are given in parentheses.}\label{fig:diag}
\end{figure}
\begin{figure}[!ht]
  \centering
  \includegraphics[width=\textwidth]{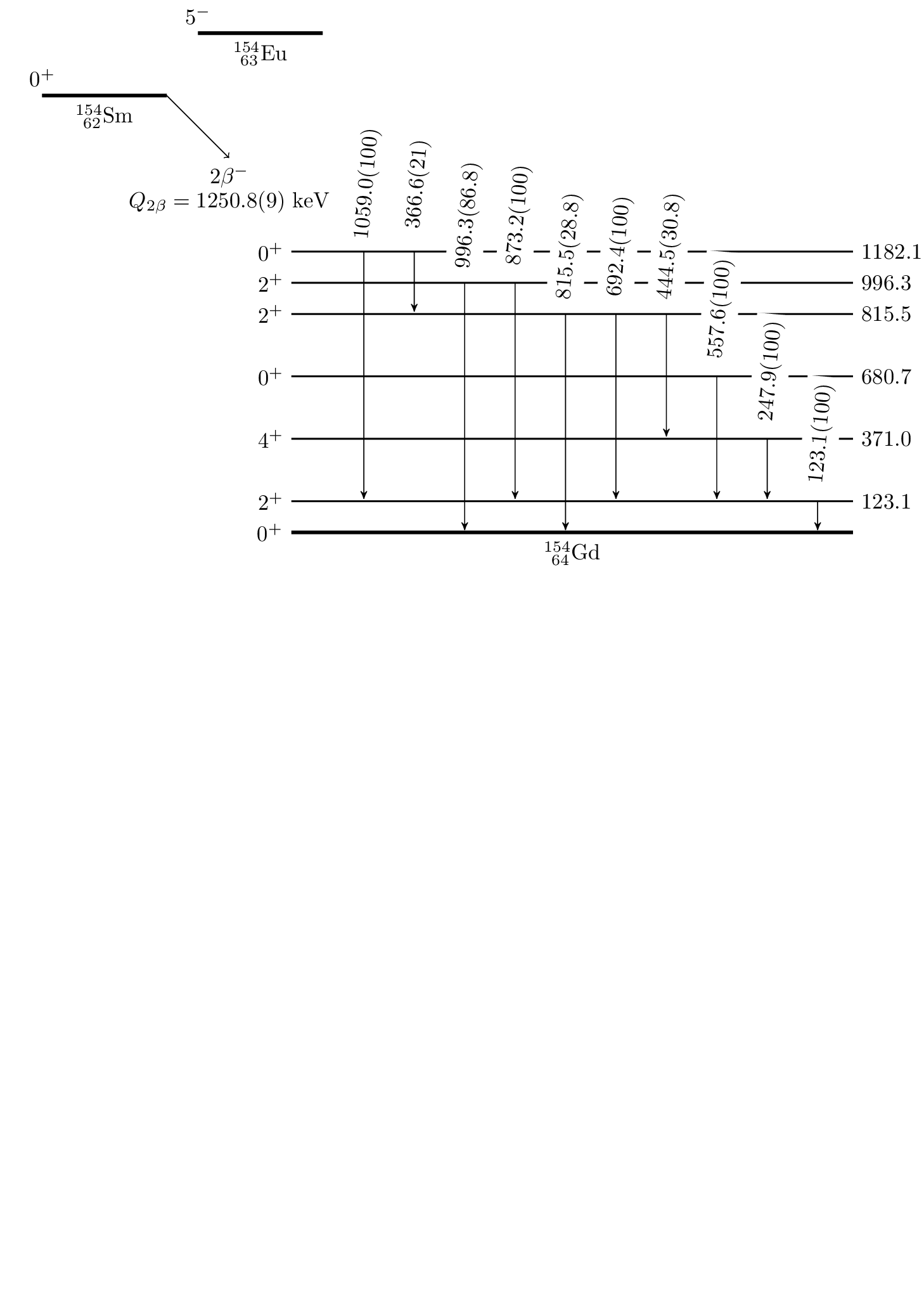}
\caption{Simplified decay scheme of $^{154}$Sm \cite{tabGd154}. The energies of the excited levels and of the emitted $\gamma$ quanta are in keV; the relative intensities of the $\gamma$ quanta
are given in parentheses. Only energy levels and $\gamma$ transitions with relative intensity larger than 3\% are shown.}\label{fig:diag154}
\end{figure}

In this paper the data collected with an ultra-low background (ULB) HP-Ge $\gamma$ spectrometer have been analyzed in order 
to obtain new limits on double beta decay processes 
in $^{144}$Sm with emission of $511$ keV $\gamma$ quanta after the 
$\beta^+$ annihilation, or $\gamma$ quanta expected in the de-excitation of the daughter nuclei.
A simplified scheme of the double beta decay of $^{144}$Sm is presented in Fig. \ref{fig:diag}.

Moreover, $^{154}$Sm $2\beta^-$ decay is possible to the ground and to the several excited levels of 
$^{154}$Gd with subsequent emission of 
$\gamma$ quanta with energies in the range ($123$ -- $1059$) keV (see Fig. \ref{fig:diag154}). The 
characteristics of $^{144}$Sm and 
$^{154}$Sm isotopes are given in Table \ref{tabqd}.

\begin{table}[!htbp]
\begin{center}
\caption{Characteristics of the samarium isotopes candidates for what concerns double beta decay.}
\begin{tabular}{c c c c}
\hline
$2\beta$ Transition                 & Q$_{2\beta}$ (keV)        & Isotopic Abundance   & Decay Channel   \\
                                    & \cite{Q144sm}          &  (\%) \cite{isotops} &  
\cite{tretyak} \\
\hline
$^{144}$Sm $\rightarrow$ $^{144}$Nd &       1782.4(8)    & 3.08(4)               & $\epsilon\beta^+$, 
$2\epsilon$ \\
$^{154}$Sm $\rightarrow$ $^{154}$Gd &       1250.8(9)  & 22.74(14)             & $2\beta^-$ \\
\hline
\end{tabular}\label{tabqd}
\end{center}
\end{table}

The measurements with Sm are part of our program to investigate the possibility to purify lanthanide elements and, 
using the purified samples, to study the results of the purification and also to look for their 2$\beta$ processes 
in low-scale experiments \cite{bellice2014,bel19,31a,er,31b}.
The interest for purification is related mainly to $^{150}$Nd and $^{160}$Gd, also 
rare-earth nuclides, which are ones of the most promising candidates for $0\nu$2$\beta$ decay searches. 
Some details on purification can be found in \cite{31c}.

To our knowledge, theoretical estimates of the half-lives for $^{144}$Sm are absent in the literature. 
The half-life for $0\nu$2$\epsilon$ mode (ground state to ground state, g.s. to g.s., 
transition) was calculated as 1.0$\times$10$^{31}$ yr \cite{31d}. The only known experimental limits were 
obtained only
recently as 8.0$\times$10$^{8}$ yr for all the decay modes and 1.0$\times$10$^{15}$ yr the $0\nu\epsilon\beta^+$ 
mode (g.s. to g.s.) \cite{Noz}. 
For $^{154}$Sm, the theoretical results for neutrinoless mode are at level of 10$^{24}$ -- 10$^{26}$ yr 
(for g.s. to g., 
and neutrino mass of 1 eV) \cite{51a,51b,51c,51d}. For the two neutrino decay 
of $^{154}$Sm (g.s. to g.s.), the theoretical results are at level of 10$^{20}$ -- 10$^{23}$ yr 
\cite{51a,51e,51f,51g,51h,51i}.
The half-life for $2\nu$2$\beta$ decay of $^{154}$Sm to the level of $^{154}$Gd (123 keV) was calculated 
as 1.4$\times$10$^{29}$ yr \cite{51a} 
(it is rather long, in particular, due to the change in spin by 2 units). Experimental limit was set only for 
transition to 2$_1^+$ level (123 keV) as 2.3$\times$10$^{18}$ yr \cite{51j}.

\section{The experiment}

\subsection{The purification of samarium oxide}

Commercially available samarium oxide Sm$_2$O$_3$ from Stanford Materials Corporation ($>$99.5\% 
total rare earth oxide (TREO) and $>$99.999\% 
of Sm$_2$O$_3$/ TREO) was used as starting material. 
The sample was firstly measured by ICP-MS and by ULB HP-Ge $\gamma$-spectroscopy to determine the initial purity. 
The main radioactive contaminants of the material are 
radium, thorium, uranium and lutetium, typical impurities in the lanthanide elements 
\cite{bel19,32a,32b,EGS,decay0}. 
 
After those 
measurements, purification procedures have been applied; they consist of several stages of chemical and physical transformations: 
(i) dissolution of the initial oxide in nitric acid; (ii) fractional precipitation of the Sm(OH)$_3$; (iii) 
liquid-liquid extraction; (iv) complete precipitation of the Sm(OH)$_3$ and 
(v) samarium oxide recovery.

To prepare homogeneous aqueous solutions of samarium, diluted nitric acid (Alfa Aesar) was added to 
the suspension of Sm$_2$O$_3$ in deionized 
water (18.2 M$\Omega \times$cm). The obtained solution was analysed and the concentrations 
determined as 1.58 M for Yb(NO$_3$)$_3$ and 2.5 M for residual HNO$_3$, respectively.

After the first step, ammonia gas was added to the acidic solution to cause partial (fractional) precipitation of lanthanides. At the same time also
impurities like Th, Fe hydroxides started to co-precipitate. This is possible since the hydroxides of thorium and iron precipitate at
a lower pH level than Sm hydroxide. After this procedure the final pH of the solution was about 6.5 and the mass of the precipitated Sm was 3.4\% 
of 
the initial mass. The obtained amorphous 
Sm(OH)$_3$ sediment was separated from the supernatant liquid by using a centrifuge, then annealed to Sm$_2$O$_3$ and analysed by ICP-MS in order to 
determine 
the efficiency of the co-precipitation of the impurities (see Table \ref{rad-pur}).
\begin{table}[!ht]
\begin{center}
\caption{Results of the ICP-MS analysis.}
\begin{tabular}{c c c c}
  \hline
 \multirow{1}*{Element}  &\multicolumn{3}{c}{Concentration (ppb)} \\
 \cline{2-4}  &	Initial material  	& Sediment after	                & Purified material\\
              &	   	                & fractional precipitation        	& \\
              &	   	                & (waste)                          	& \\
  \hline
    K  &	$<$2000	&1639	&424\\
    Fe	&   2400    &27441	&1190\\
    Th	&   2.2	    &9.5	&$<$0.3\\
    U	&   48	    &321	&$<$0.1\\
    Pb	&   90	    &466	&56\\
    La	&   1300	&700	&1200\\
    Lu	&   6200	&12700	&4100\\
  \hline
\end{tabular}\label{rad-pur}
\end{center}
\end{table}
This part of samarium oxide was excluded from the further purification procedure.

The liquid-liquid extraction procedure was applied for the purification of the aqueous solution from uranium and 
thorium traces. In this stage the solution was acidified with diluted nitric acid down to pH equal to 1. 
The tri-n-octylphosphine oxide (TOPO, 99\%, Acros Organics) was used as organic complexing agent for 
binding U and Th, while toluene ($\geq$ 99.7\%, ACS reagent, Sigma-Aldrich) was used as solvent liquor. 
Considering the very low concentration, the concentration of TOPO 
in toluene did not exceed 0.1 mol/l. Two immiscible liquids (aqueous solution and organic solution) 
were placed into a separation funnel in volumetric ratio 1:1 and shaken up for 
a few minutes. Uranium and thorium interact with TOPO forming 
organo-metallic complexes that have much higher solubility in the organic phase than in the water solution. 
This leads to extraction of U and Th into the organic liquid.
After the separation of the purified aqueous solution, samarium was completely precipitated in form of hydroxide
by adding ammonia gas. The obtained sediments were separated, dried and annealed at 900$^\circ$C for 
a few hours. Finally, 342 g of purified Sm oxide were obtained.

\subsection{The low counting experiment}\label{p:resol}

The experiment  was carried out at the STELLA facility of the LNGS by using the ULB HP-Ge detector ``GeCris'' with volume 465 cm$^3$. The detector is 
shielded with low radioactivity lead 
($\sim$ 25 
cm), copper ($\sim$ 5 cm), and, in  the inner-most part, with archaeological Roman lead ($\sim$ 2.5 cm). The set-up is placed in an air-tight 
poly-methyl-methacrylate box and flushed with
 high purity nitrogen gas to exclude the environmental
radon. The purified sample of Sm$_2$O$_3$ with mass 342 g was enclosed in
a cylindrical  polystyrene box on the HP-Ge detector end cap.

The energy resolution of the detector was estimated by using background
$\gamma$-ray peaks with energies 238.6 keV ($^{212}$Pb), 338.3 keV ($^{228}$Ac), 463.0 keV ($^{228}$Ac), 583.2 keV
($^{208}$Tl), 661.7 ($^{137}$Cs), 727.3 keV ($^{212}$Bi), 911.2 keV ($^{228}$Ac), 1460.8 keV ($^{40}$K) and
2614.5 keV ($^{208}$Tl), measured in the contiguous experiment 
of Ref. \cite{bellice2014} with a cerium oxide sample, and verified with the present Sm$_2$O$_3$ data taking.

\begin{figure}[!ht]
  \centering
  \includegraphics[width=\textwidth]{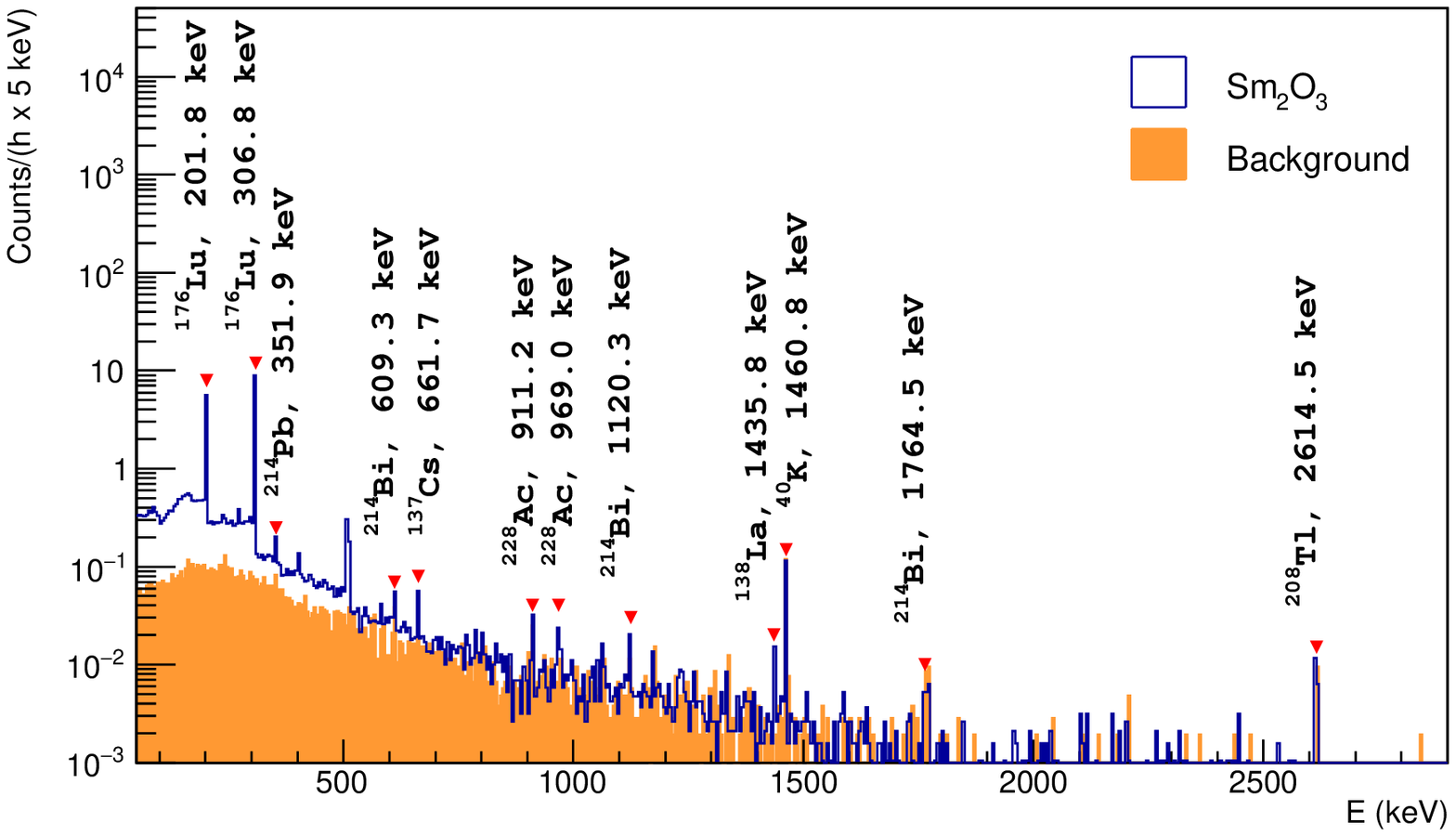}
  \caption{(Color online) Energy spectra measured by the ULB 
HP-Ge $\gamma$ spectrometer with the Sm$_2$O$_3$ sample over 1899 h (Sm$_2$O$_3$)
and without the sample over 1046 h (Background).}\label{fig:spectra}
\end{figure}

The energy resolution of the detector depends on the energy of the $\gamma$ quanta, $E_\gamma$, according to: 
FWHM(keV) $ = \sqrt{1.41 + 1.97\times10^{-3} E_\gamma}$, where $E_\gamma$ is in keV.
The data with the Sm$_2$O$_3$ sample were taken over 1899 h, while
the background spectrum was collected for 1046 h.
The two spectra, normalized to the time of measurements, are presented in Fig. \ref{fig:spectra}.
\begin{table}[!htpb]
\begin{center}
\caption{Radioactive contaminations of the samarium oxide before and after applying the purification procedures as measured with the ultra-low background HP-Ge $\gamma$ spectrometer.
The upper limits are given at 95\% confidence level (C.L.), and the uncertainties are $1 \sigma$.}
\begin{tabular}{l c c c c r}
\hline
 \multirow{2}*{~} &\multirow{2}*{Chain} & \multirow{2}*{Nuclide} &\multicolumn{2}{c}{Activity 
(mBq/kg)} &~\\
 \cline{4-5} &         &             & Before purification            & After purification &~\\
 \hline
~ &~          & $^{40}$K              & $< 11$                        &   $11\pm5$          &~          \\
~ &~          & $^{137}$Cs            & $< 0.67$                      &   $0.8\pm0.2$       &~          \\
~ &~          & $^{138}$La            & -                             &   $0.81\pm0.16$               &~\\
~ &~          & $^{152}$Eu            & -                             &   $< 0.50$                    &~\\
~ &~          & $^{154}$Eu            & -                             &   $< 0.43$                    &~\\
~ &~          & $^{176}$Lu            & ($0.23\pm0.02$) $\times 10^{3}$&   $(0.203\pm0.015)\times10^3$  &~\\
\cmidrule{2-5}
~ &$^{232}$Th          & $^{228}$Ra            & $11\pm2$                      &   $1.4\pm0.7$        &~\\
~ &~          & $^{228}$Th            & $24\pm3$                      &   $< 1.1$                     &~\\
\cmidrule{2-5}
~& $^{238}$U  & $^{234}$Th            & $< 1.7\times 10^3$            &   $< 0.13\times10^3$          &~\\
~& ~          & $^{234m}$Pa           & $(0.22\pm0.05$)$\times10^3$   &   $< 18$                      &~\\
~& ~          & $^{226}$Ra            & $6\pm1$                       &   $1.5\pm0.4$                 &~\\
\cmidrule{2-5}
~& $^{235}$U  & $^{235}$U             & $31\pm7$                      &   $< 2.7$                     &~\\
~&~           & $^{231}$Pa            & -                             &   $< 15$                      &~\\
~&~           & $^{227}$Th            & -                             &   $5\pm2$                     &~\\
~& ~          & $^{223}$Ra            & -                             &   $< 10$                      &~\\
~& ~          & $^{211}$Pb            & -                             &   $< 6.7$                     &~\\
~& ~          & $^{207}$Tl            & -                             &   $< 52$                      &~\\
 \hline
\end{tabular}\label{rad-cont}
\end{center}
\end{table}
Some excess of $^{137}$Cs, $^{138}$La, $^{176}$Lu and $^{214}$Bi (daughter of $^{226}$Ra) with respect to the
background data was observed; this allowed us to estimate the concentration of these radionuclides in the sample.
The activities of the nuclides in the Sm$_2$O$_3$ sample after its purification are also shown in Table \ref{rad-cont}.
The concentration of $^{176}$Lu remained almost the same as prior to the
purification\footnote{It is similar to what was observed in the purification of other rare earth 
elements, e.g. \cite{er,bel19,31a}.}; this is due 
to the high chemical affinity between the Sm and Lu, both lanthanides, while the activity of 
$^{226}$Ra is
decreased by a factor 4. The activities of $^{228}$Th, $^{228}$Ra, $^{234m}$Pa and 
$^{235}$U were reduced 
by about one order of magnitude (see Table \ref{rad-cont}).

\subsection{The search for double beta decay processes}

There are no peculiarities in the energy spectrum of the
Sm$_2$O$_3$ sample that could be identified as double beta decay of the Sm
isotopes. Therefore, half-life
limits of the $\epsilon\beta^+$ and $2\epsilon$ processes in $^{144}$Sm 
and of the $2\beta^-$ processes in $^{154}$Sm were estimated. The lower half-life limits 
were calculated using the following equation: \begin{eqnarray}
T_{1/2,lim} = N\times \eta \times t \times \frac{\ln2}{S_{lim}}\mbox{,}
\end{eqnarray}
where: i) $N$ is the number of nuclei of interest in the sample; ii) 
$\eta$ is the full energy peak (FEP) detection
efficiency of $\gamma$ quanta (including the $\gamma$-ray emission intensity); 
iii) $t$ is the time of measurement; iv) $S_{lim}$ is the upper limit on the
number of events of the effect, searched for, that can be excluded at a given
C.L.. In the present work all the $S_{lim}$ values and, therefore,
the half-life limits are given at 90\% C.L. The detection efficiencies of 
the 
processes searched for were Monte-Carlo simulated by using the EGSnrc 
\cite{EGS} package with initial kinematics given by the DECAY0 event 
generator \cite{decay0}.
 The Sm$_2$O$_3$ sample contained $2.69\times10^{23}$ and $3.64\times10^{22}$ nuclei of $^{154}$Sm and $^{144}$Sm, respectively.

\subsection{The search for the $\epsilon\beta^+$ and the $2\epsilon$ processes in $^{144}$Sm }

In the case of $2\epsilon$ decays, a cascade of X-rays and Auger electrons is expected.
Moreover, the double electron capture in $^{144}$Sm is also allowed to excited levels of $^{144}$Nd with 
subsequent emission of gamma quanta that can be detected with the HP-Ge spectrometer.

 In order to estimate the limits on the $2\nu2\epsilon$ and $0\nu2\epsilon$ decays
of $^{144}$Sm to the $2^+$ excited levels of $^{144}$Nd, the energy spectrum
of the Sm$_{2}$O$_{3}$ sample was fitted in the energy intervals where intense $\gamma$-ray peaks
from the de-excitation process are expected. The limits obtained for the double electron
capture of $^{144}$Sm to the excited levels of $^{144}$Nd are given in Table \ref{HLlim}.

For the $0\nu2\epsilon$ decay mode of $^{144}$Sm to the ground and the excited levels
of $^{144}$Nd 
only the captures from $K$ and $L$ shells were considered here;
we assume that the energy excess is taken away by bremsstrahlung $\gamma$ quanta 
with energy $E_\gamma = Q_{2\beta} - E_{b1} - E_{b2} - E_{exc}$,
where $E_{bi}$ are the binding energies 
\begin{table}[!ht]
\begin{center}
\caption{The half-life limits on the $2\beta$ processes in $^{144}$Sm and $^{154}$Sm.}
\resizebox{\textwidth}{!}{
\begin{tabular}{l c c c c c c c}
\hline
    Process             &   Decay     & Level of      & E$_\gamma$  & FEP detection    & $S_{lim}$ &   Experimental \\
    of decay            &  mode       &daughter       & (keV)       & efficiency   &           & limit (yr)    \\
                        &             &nucleus        &             &(\%)          &           & at 90\% C.L.  \\
                        &             &(keV)          &             &              &           &               \\
 \hline
    $^{144}$Sm $\rightarrow$ $^{144}$Nd & & & & & & \\
    2$K$                  &2$\nu$       & $2^+$ 696.6   & 696.5       &      2.93     &  10     &  $\geq1.6\times10^{19}$  \\
    2$K$                  &2$\nu$       & $2^+$ 1560.9  & 864.3         &      2.14     & ~ ~0.90   & $\geq1.3\times10^{20}$  \\
    $\epsilon\beta^+$     &2$\nu$       & g.s.          & 511.0         &      6.55     &  10~   & $\geq3.6\times10^{19}$  \\
    $\epsilon\beta^+$   &2$\nu$         & $2^+$ 696.6     & 511.0         &     5.76      &  10~   & $\geq3.2\times10^{19}$  \\
                        &               &               &             &               &           &                       \\
    2$K$                  &0$\nu$       & g.s.          & 1695.3~       
&      1.87     &  ~2.3   &  $\geq4.4\times10^{19}$  \\
    $KL$                  &0$\nu$       & g.s.         & 1731.7         &      
1.84     &  ~6.0   &  $\geq1.7\times10^{19}$  \\
    2$L$      &0$\nu$     & g.s.        & 1768.2         &      1.81     & 
~ 7.0  &  $\geq1.4\times10^{19}$  \\
    2$K$                  &0$\nu$       & $2^+$ 696.6     &  696.5        &      2.57     &   10~    &  $\geq1.4\times10^{19}$  \\
    2$K$                  &0$\nu$       & $2^+$ 1560.9   &  864.3        &      2.08     &   ~ ~0.90    &  $\geq1.3\times10^{20}$  \\
    $\epsilon\beta^+$   &0$\nu$         & g.s.          &  511.0        &      6.38     &   10~  &  $\geq3.5\times10^{19}$  \\
    $\epsilon\beta^+$   &0$\nu$         & $2^+$ 696.6     &  511.0        &      5.77     &   10~  &  $\geq3.2\times10^{19}$  \\
 \hline
    $^{154}$Sm $\rightarrow$ $^{154}$Gd & & & & & & \\
    $2\beta^-$          &2$\nu$+0$\nu$& $2^+$ ~123.1   &    ~123.1     &      0.40     &    27~    &  $\geq6.0\times10^{18}$  \\
    $2\beta^-$          &2$\nu$+0$\nu$& $0^+$ ~680.7   &    ~557.6     &      3.08     &    ~ 4.8 &  $\geq2.6\times10^{20}$  \\
    $2\beta^-$          &2$\nu$+0$\nu$& $2^+$ ~815.5   &    ~692.4     &      1.80     &    ~ 2.4 &  $\geq3.0\times10^{20}$  \\
    $2\beta^-$          &2$\nu$+0$\nu$& $2^+$ ~996.3   &    ~873.2     &      1.39     &    ~ 2.8 &  $\geq2.0\times10^{20}$  \\
    $2\beta^-$          &2$\nu$+0$\nu$& $0^+$ 1182.1   &    1059.0    &      1.99     &    ~ 1.0 &  $\geq8.0\times10^{20}$  \\
 \hline
\end{tabular}\label{HLlim}
}
\end{center}
\end{table}
of the captured electrons in the atomic shells of the
daughter neodymium atom, and $E_{exc}$ is the energy of the excited level.

In the case of $2\nu2K$ or $0\nu2K$ capture in $^{144}$Sm to the first ($2^+$, 696.6 keV) excited level,
$\gamma$ quanta with 696.5 keV energy are expected.
The $S_{lim}$ value was obtained by fitting the experimental data in the energy interval where the peak is expected, 
considering a linear background model plus a Gaussian peak 
at 696.5 keV energy, taking into account the energy resolution described above. In the range (670--730) keV 
the fit provides 
the area of the peak searched for with $\chi^2/n.d.f.\simeq0.99$ (n.d.f. = number of degrees of freedom); it is: 
($3.1\pm4.3$) counts, that is no evidence on the effect searched for.  Thus, the $S_{lim}$ value
was estimated using the procedure of Ref. \cite{FelCou}; $S_{lim}$ = 10 counts was obtained. The FEP detection 
efficiencies for the $2\nu2K$ and $0\nu2K$ modes, simulated by a dedicated Monte-Carlo code, were: $\eta=2.93\%$, $\eta=2.57\%$, respectively.
Taking into account the number of $^{144}$Sm nuclei in the sample, the half-life limits of Table \ref{HLlim} were obtained.

A similar procedure was followed to estimate the $S_{lim}$ value for the 864.3 keV peak expected for the $2\nu2K$ and $0\nu2K$ transitions to 
the $^{144}$Nd 1560.9 keV level. The fit, performed within the energy range (855--885) keV 
($\chi^2/n.d.f.$ $\simeq 0.99$), gives ($-1.1\pm1.1$) counts; 
the $S_{lim}$ value, estimated by applying the procedure of Ref. \cite{FelCou}, was: 0.90 counts. 
The FEP detection efficiencies, simulated by Monte-Carlo code, are $\eta=2.14\%$ and $2.08\%$, 
respectively. The derived half-life limit in both cases is:  $T_{1/2}\geq 1.3\times10^{20}$ yr.

In the case of $0\nu2K$ capture in $^{144}$Sm to the $^{144}$Nd g.s., 1695.3 keV $\gamma$ 
quanta are expected.
Following the same procedure described above, the fit provides -- in the (1670--1720) keV energy range 
($\chi^2/n.d.f.\simeq0.30$) -- 
an area for the $0\nu2K$ process equal to ($0.0\pm1.4$) counts.  The $S_{lim}$ value, estimated using 
the procedure of Ref. \cite{FelCou}, was 2.3 counts. The FEP detection efficiency, simulated by Monte-Carlo code, was: 
$\eta=1.87 \%$. The half-life limit, obtained for this process, is $T_{1/2}\geq 4.4\times10^{19}$ yr at 90\% C.L.

\begin{figure}[!ht]
\centering
\includegraphics[width=\textwidth]{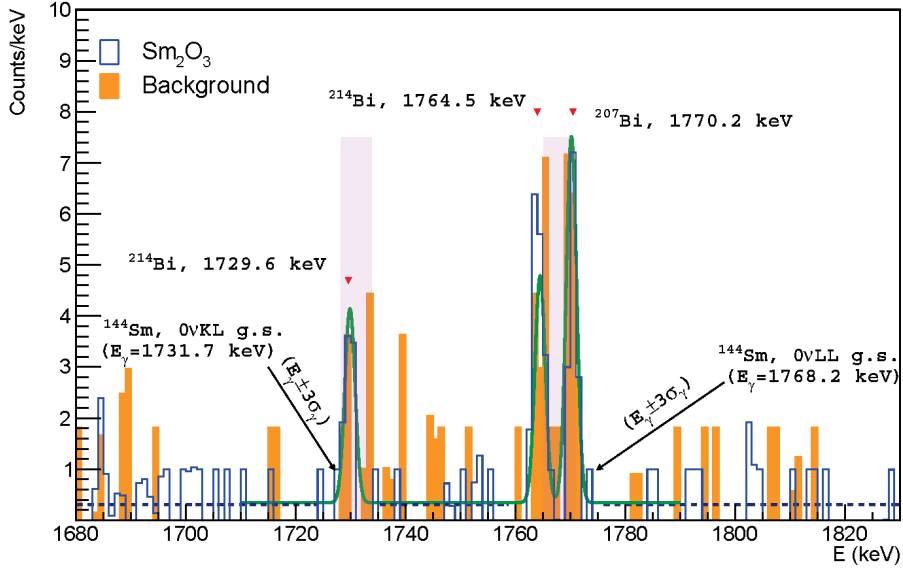}
\vspace{-0.8cm}
\caption{Part of the energy spectrum measured with the
Sm$_2$O$_3$ sample over 1899 h, where the $\gamma$ peaks from the $0\nu KL$ and $0\nu 2L$ processes in $^{144}$Sm to the g.s. of $^{144}$Nd are expected. 
The fit (green, color on-line) of the background data (a straight line plus the 1729.6 keV and 1764.5 keV Gaussian $\gamma$ peaks from $^{214}$Bi and 
1770.2 keV peak from $^{207}$Bi) is shown.
The constant fit (dashed blue on-line) of the data with the Sm$_2$O$_3$  $6\sigma$ far from the mentioned $\gamma$-ray 
peaks of $^{214}$Bi and  $^{207}$Bi is also reported.
The boxes (transparent pink on-line) indicate the energy range ($E_{\gamma}\pm 3\sigma_{\gamma}$) expected for the $0\nu KL$ and $0\nu 2L$ decay processes to the g.s. of $^{144}$Nd 
in the energy spectra measured with the Sm$_2$O$_3$ sample (see text).}\label{fig:fit00}
\end{figure}

In case of $0\nu KL$ and $0\nu 2L$ processes to the g.s. of $^{144}$Nd 
(see Fig. \ref{fig:fit00}), the energies of the $\gamma$ quanta are 1731.7 
keV and 1768.2 keV 
(FEP efficiencies: $\eta_{1731.7}=1.84\%$ and $\eta_{1768.2}=1.81\%$), 
respectively.

\begin{figure}[!ht]
  \centering
\vspace{-0.8cm}
   \includegraphics[width=1.\textwidth]{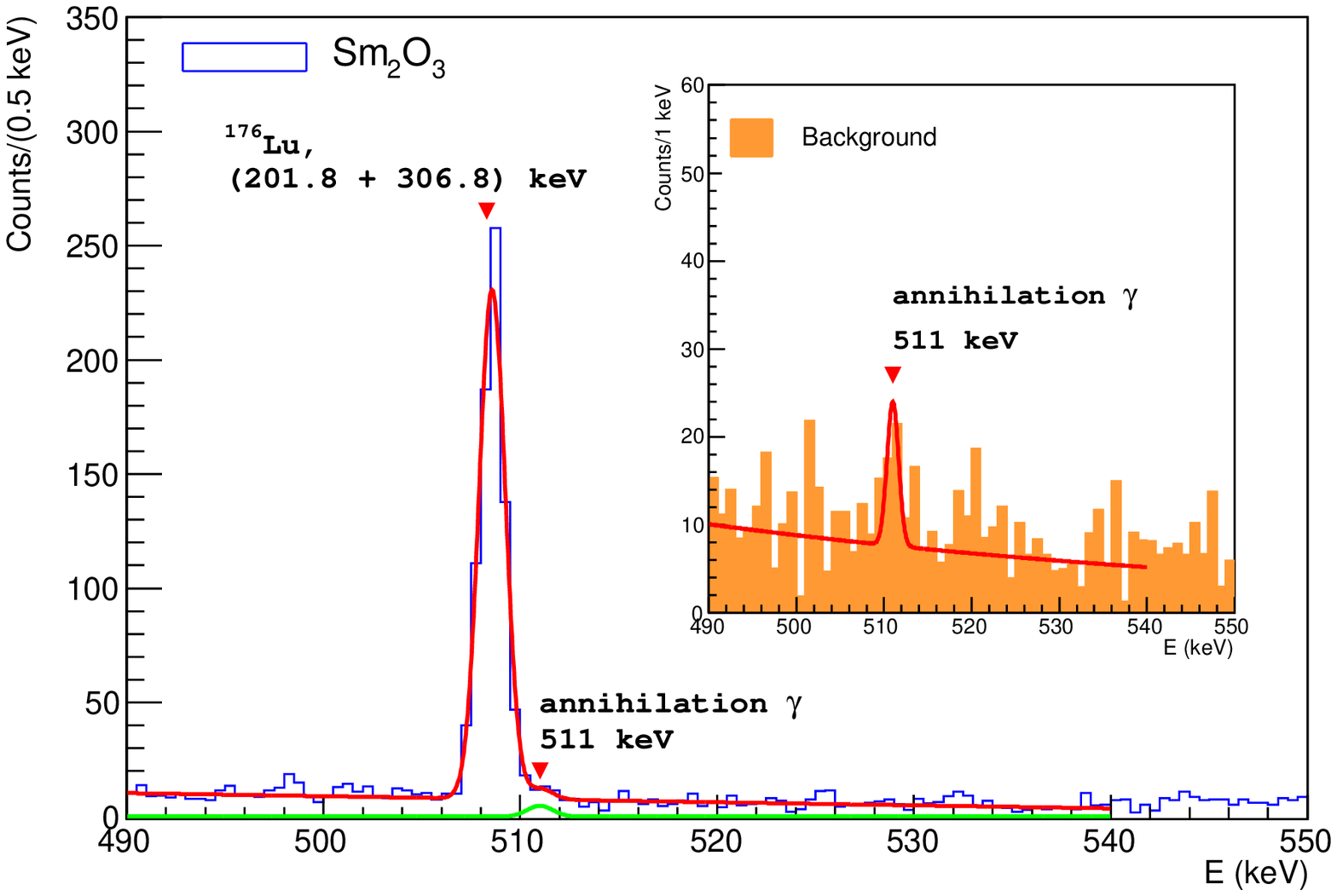}
\vspace{-0.8cm}
  \caption{Energy spectrum measured with the samarium oxide sample over 1899 h and
background spectrum (inset frame) measured over 1046 h in the vicinity of
the 511 keV annihilation peak (the red -- colored on-line -- line in the inset frame is the fit
of the background data). In the Sm$_2$O$_3$ sample the 508.6 keV peak belonging to $^{176}$Lu is also present.
The fit of the Sm$_2$O$_3$ sample data takes into account the 511 keV
$\gamma$ quanta (green on-line) and the 508.6 keV peak of 
$^{176}$Lu. The red (colored on-line) line in the main figure is the total fit
of Sm$_2$O$_3$ sample (see text).}\label{fig:fit511}
\end{figure}
As shown in Fig. \ref{fig:fit00}, in such cases the $\gamma$-ray peaks of $^{214}$Bi and $^{207}$Bi are present -- 
in the energy range of interest -- in both 
energy spectra measured with and without the Sm$_2$O$_3$ sample. 
However, comparing the counts of each $\gamma$-ray peaks measured in the two spectra, we can note that the relative residuals: $(3.1\pm3.6)$, 
$(-1.1\pm5.7)$ 
and $(-4.1\pm5.1)$ counts for the lines 
at 1729.6 keV, 1764.5 keV and 1770.2 keV, respectively, are compatible with zero.
Therefore, we can be confident that the $^{214}$Bi and  $^{207}$Bi $\gamma$-ray peaks in the energy spectra 
with 
the Sm$_2$O$_3$ sample are due to the intrinsic background of the experimental setup.
Thus, the background model for the energy range of the expected peaks is made of a flat component and of the $^{214}$Bi 
and $^{207}$Bi $\gamma$ peaks; in particular: (i) the contribution of the three peaks is determined from the background 
data by using a flat component and three Gaussian functions (green on-line in the Fig. \ref{fig:fit00}); (ii) the flat 
component in the samarium spectrum is evaluated by fitting of the samarium spectrum with a straight line, 
excluding 6$\sigma$ from the $^{214}$Bi and $^{207}$Bi peaks (dashed blue on-line in Fig. \ref{fig:fit00}). The energy 
range of interest, used for this analysis, is $E_{\gamma}\pm 3\sigma_{\gamma}$, where $E_{\gamma}$ are the energy of the $\gamma$ quanta 
expected for the $0\nu KL$ and $0\nu 2L$ decays to the g.s. of $^{144}$Nd (pink transparent boxes on-line  
in Fig. \ref{fig:fit00}). Here, using the described background model, one can estimate the residual counts for the $0\nu KL$ 
and $0\nu 2L$  decays to the g.s. of $^{144}$Nd: $(-1.3 \pm 4.4)$ and $(-2.4 \pm 5.6)$ counts, respectively. 
According to the  procedure of Ref. \cite{FelCou}, the derived $S_{lim}$ values are 6.0 counts and 7.0 
counts, respectively. Thus, the following half-life limits are obtained: 
$T_{1/2}\geq 1.7\times10^{19}$ yr and $T_{1/2}\geq 1.4\times10^{19}$ yr, respectively.


One positron is expected in the $2\nu\epsilon\beta^+$ ($0\nu\epsilon\beta^+$) decay of $^{144}$Sm with an energy up to
$761$ keV. The
annihilation of the positron should produce two 511 keV $\gamma$ quanta resulting in an extra
counting rate in the annihilation peak.
To estimate the $S_{lim}$ value, the energy spectra accumulated with the
Sm$_2$O$_3$ sample and the background data were fitted in the energy interval (490--540) keV; the fit of 
the 
Sm$_2$O$_3$ sample data takes into account also the 508.6 keV peak of $^{176}$Lu (see
Fig. \ref{fig:fit511} ). Considering the area of the annihilation peak in the background, in the 
data accumulated with the
samarium oxide sample there are ($-11\pm12$) residual events in the 511 keV peak.
Since there is no evidence of the effect searched for, we derive $S_{lim}$ = 10 counts.
The FEP efficiencies for the $0\nu \epsilon \beta^+$ and $2\nu \epsilon \beta^+$ processes 
are very similar (see Table  \ref{HLlim}). The obtained half-life limits for the transition 
to the g.s are $3.5 \times 10^{19}$ yr and $3.6 \times 10^{19}$ yr, respectively, 
while for the transition to the $2^+$ 696.6 keV level of $^{144}$Nd the half-life limit $3.2 \times 
10^{19}$ yr is obtained in both cases.

\subsection{Search for $2\beta^-$ processes in $^{154}$Sm }

The double beta decay of $^{154}$Sm is possible to the ground state and to the excited levels
of $^{154}$Gd with energy  123.1 keV, 680.7 keV, 815.5 keV, 996.3 keV 
and 1182.1 keV (see Fig.  \ref{fig:diag154}).
Our experiment is sensitive only to the transition to excited levels.

\begin{figure}[!ht]
  \centering
  \includegraphics[width=\textwidth]{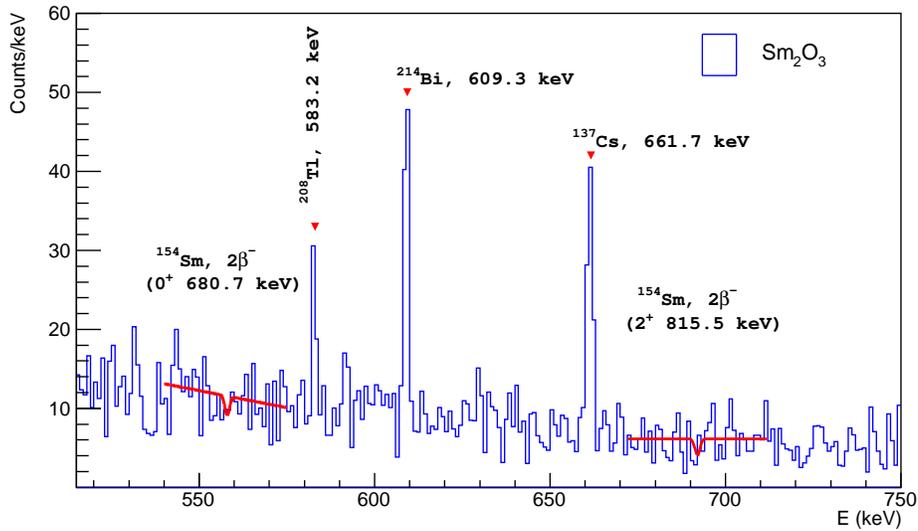}
\vspace{-0.2cm}
  \caption{Part of the energy spectrum measured with the
Sm$_2$O$_3$ sample over 1899 h, where the $\gamma$-ray peaks from the $2\beta^-$ in $^{154}$Sm to the $0^+$ 680.7 keV and 
$2^+$ 815.5 keV excited levels of the $^{154}$Gd are expected. The fits are shown by solid (red on-line) lines; 
the energy ranges of the fits are (540--575) keV and (672--712) keV.}
\label{fig:fit03}
\end{figure}

\begin{figure}[!ht]
  \centering
  \includegraphics[width=\textwidth]{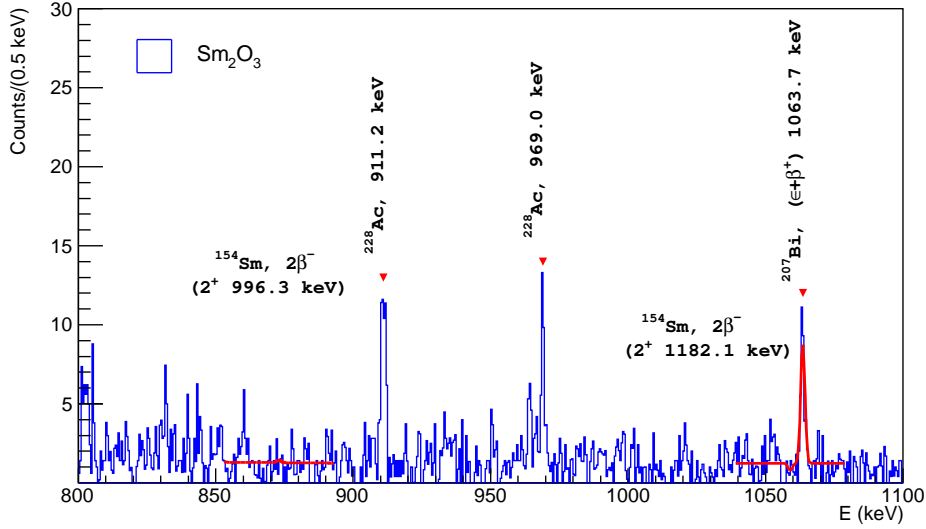}
  \vspace{0.2cm}
  \caption{Part of the energy spectrum accumulated with the
Sm$_2$O$_3$ sample over 1899 h, where the $\gamma$-ray peaks from the $2\beta^-$ in $^{154}$Sm to $2^+$ 996.3 keV and $0^+$ 1182.1 keV excited levels of 
$^{154}$Gd are expected. The fits are shown by solid (red on-line) lines. The energy range of the fits are (853--893) keV and (1039--1079) 
keV.}
\label{fig:fit04}
\end{figure}

As regards the transition to the first excited level at 123.1 keV, the energy spectrum acquired with the samarium oxide sample was fitted in
the energy interval (110--135) keV by a model consisting of a Gaussian function centered  at
123.1 keV (to describe the effect searched for) and a straight line as a background model. The
fit gives an area of the 123.1 keV peak equal to (11 $\pm$ 10) counts; thus, there is no evidence
for the effect searched for. Therefore, according to Ref. \cite{FelCou}, we derive $S_{lim}$ = 27 counts. Taking
into account the number of $^{154}$Sm nuclei in the sample, and the FEP detection efficiency
$\eta=0.40\%$, we have obtained the limit on the $2\beta^-$ decay of $^{154}$Sm to the first $2^{+}$ excited level of
$^{154}$Gd: $T_{1/2}\geq 6.0\times10^{18}$ yr. This limit is for the sum of the 
$2\nu2\beta^-$ and $0\nu2\beta^-$ modes, since they
cannot be distinguished with the $\gamma$-spectrometry method.

As regards the other excited  levels of the $^{154}$Gd, 680.7 keV, 815.5 keV, 996.3 keV, 1182.1 keV de-excitation gamma peaks are  expected at 557.6 
keV, 692.4 keV, 873.2 keV and 1059.0 keV, respectively (see Fig. \ref{fig:diag154}). The spectrum in the vicinity of the
expected peaks is shown in Figs. \ref{fig:fit03} and \ref{fig:fit04}.
In the first three cases the energy spectrum with the Sm$_2$O$_3$ sample was
fitted by the sum of a Gaussian function (to describe the peak searched for) and of a straight line
to describe the background.

In the case of the $2\beta^-$ decay to the 1182.1 keV level of $^{154}$Nd, also the 1063.7 keV peak of $^{207}$Bi was included in the fit to 
approximate the background in a large enough energy interval around the peak searched for (see Fig. \ref{fig:fit04}).
To perform the fits, the energy range was fixed for these 4 cases as: (540--575) keV, (672--712) keV, (853--893) keV and
(1039--1079) keV, respectively. The results of the fits give for the effects searched for the values: $(-4.1\pm5.1)$, 
$(-3.7\pm3.3)$, $(0.3\pm1.5)$, $(-0.9\pm1.1)$ counts, respectively. According to the procedure given in Ref. \cite{FelCou}, the 
corresponding S$_{lim}$ values are 4.8, 2.4, 2.8 and 1.0 counts, respectively; 
thus, the half-life limits are in the range $T_{1/2}\geq (2.0 - 8.0)\times10^{20}$ yr (see Table  \ref{HLlim}).

\section{Conclusions}

A method of samarium purification from radioactive contaminants in traces, based on
the liquid-liquid extraction was performed. 
In particular, traces of $^{40}$K, $^{137}$Cs and $^{226}$Ra were observed in the purified Sm$_2$O$_3$ sample at the level of 
$\sim$(1 - 10) mBq/kg, while other contaminants --
as e.g. $^{228}$Th -- are below the experimental sensitivity of $\sim$ 1 mBq/kg. Instead the adopted protocol of samarium purification 
is not effective for the segregation of $^{176}$Lu. Further investigations are in progress.

The $2\epsilon$ and $\epsilon\beta^+$ decay modes
in $^{144}$Sm, and the $2\beta^-$ of $^{154}$Sm to the excited levels $2^+$ or $0^+$ 
of $^{154}$Gd were searched for using 342 g of a highly purified Sm$_2$O$_3$ sample and the
ultra-low background HP-Ge $\gamma$ spectrometer at the
STELLA facility of the Gran Sasso National Laboratory. For the first
time, limits on the different decay modes and channels of double beta decay of $^{154}$Sm and $^{144}$Sm
were set at a level of $T_{1/2}$ $> 10^{18}-10^{20}$ yr. This sensitivity is not so far from 
those of the most sensitive $2\beta^+$ experiments, which are typically  at a level of $T_{1/2}$ $\sim 10^{21} - 10^{22}$ yr 
\cite{Baindi,Baindi1,Krindi,Krindi1,bbpositivi,bbpositivi1,bbpositivi2,bbpositivi3,bbpositivi4,bbpositivi5,bbpositivi6,bbpositivi7,bbpositivi8,reso3,Krivo}.
The  $T_{1/2}$ limits, obtained in this work, for the $^{144}$Sm $2\epsilon$ and $\epsilon\beta^+$ processes are 10 -- 11 orders of magnitude higher
than those recently derived in Ref. \cite{Noz}.

\section{Acknowledgments}
The group from the Institute for Nuclear Research (Kyiv, Ukraine) was supported in part by the program of the National Academy of Sciences 
of Ukraine ``Fundamental research on high-energy physics and nuclear physics'' (international cooperation)”.
O.G.P. was supported in part by the project ``Investigations of
rare nuclear processes'' of the program of the National Academy of
Sciences of Ukraine ``Laboratory of young scientists'' (Grant No.
0118U002328).

\end{document}